\documentclass[11pt,twoside]{article}
\usepackage{CAGN2019}
\usepackage{graphicx}
\usepackage[T1]{fontenc} 
\usepackage{latexsym}
\usepackage{verbatim}
\usepackage{ifpdf}  
\usepackage{amsmath,amssymb}
\ifpdf  
      \DeclareGraphicsExtensions{.pdf,.png,.jpg}  
\else  
      \DeclareGraphicsExtensions{.eps}  
\fi 
\newcommand\ion[2]{#1$\;${\scriptsize\rmfamily{#2}}\relax}

\begin{document}

\vskip 1.0cm
\markboth{M. Valerdi et al.}{Primordial Helium Abundance}
\pagestyle{myheadings}
%
%
\vspace*{0.5cm}
\parindent 0pt{Contributed  Paper}

\vspace*{0.5cm}
\title{A new determination of the primordial helium abundance based on the HII region NGC~346}

\author{M.~Valerdi \& A.~Peimbert}
\affil{$^1$Instituto de Astronom\'ia, Universidad Nacional Aut\'onoma de M\'exico, Apdo. Postal 70-264 Ciudad Universitaria, M\'exico}

\begin{abstract}
To understand the Universe a highly accurate ($\sim 1\%$) primordial helium abundance ($Y_{\rm P}$) determination plays an important role; it is extremely important to constrain: Big Bang Nucleosynthesis models, elementary particle physics, and the study of galactic chemical evolution. Low-metallicity \ion{H}{II} regions have been used to estimate it since their statistical uncertainties are relatively small. We present a new determination of the primordial helium abundance, based on long-slit spectra of the \ion{H}{II} region NGC~346 in the small Magellanic cloud. We found that for NGC~346: $X=0.7465$, $Y=0.2505$ and $Z=0.0030$. By assuming $\Delta Y / \Delta O = 3.3\pm0.7$ we found that the primordial helium abundance is $Y_{\rm P}= 0.2451 \pm 0.0026$ (1$\sigma$).


\end{abstract}

\section{Introduction}

According to the Big Bang model, the universe expanded rapidly from a highly compressed primordial state, which resulted in a significant decrease in density and temperature. After a few seconds, the universe cooled enough to allow the formation of certain nuclei. The Standard Big Bang Nucleosynthesis (SBBN) theory predicts that definite amounts of hydrogen, helium, and lithium were produced. If the BB was not standard some variations on these abundances is expected. Primordial nucleosynthesis is believed by most cosmologists to have taken place in the interval from roughly $10$s to $20$m after the Big Bang. Most of the helium as the isotope $^4$He was formed, along with small amounts of the hydrogen isotope $^2$H or D, the isotope $^3$He, and a very small amount of the isotope $^7$Li. Determination of $Y_{\rm P}$ imposes constraints the BB model on galaxy evolution models, and shows the composition of the universe when it was less that 1 Gyr old. $^4$He is very stable, and neither decays nor combines easily to form heavier nuclei.

Thus, one seeks astrophysical objects with low metal abundances, in order to measure light element abundances that are closer to primordial. Low-metallicity \ion{H}{II} regions have been used to estimate it since they are very bright and emit the relevant lines in the visual region of the spectrum. The first determination of $Y_{\rm P}$ based on observations of \ion{H}{II} regions was carried out by \citet{Peimbert1974} and several dozens have followed. Using several low-metallicity \ion{H}{II} regions has permitted to estimate the primordial helium abundance with very small statistical errors ($\lesssim$ 1\%) \citep[e.g.][]{Izotov2014, Aver2015,Peimbert2016}; nevertheless, due to the numerous systematic uncertainties, obtaining better than 1\% precision for individual objects remains a challenge.

The Small Magellanic Cloud hosts the \ion{H}{II} region NGC~346, arguably the best \ion{H}{II} region to determine the primordial helium abundance. Its low metallicity $Z\sim0.003$ ($\sim20$\% of solar) requires a relatively small extrapolation to obtain $Y_{\rm P}$. NGC~346 has many advantages with respect to other \ion{H}{II} regions: the underlying absorption correction for the helium lines can be reduced by at least an order of magnitude, the determination of the ICF(He) can be estimated by observing different lines of sight of a given \ion{H}{II} region, the accuracy of the determination can be estimated by comparing the results derived from different points in a given \ion{H}{II} region, overall there is no \ion{H}{II} region that is simultaneously at least as close, as large, and with a metallicity as low as NGC~346.

To obtain the value for $Y_{\rm P}$ it is necessary to use the correlation between the helium mass fraction $Y$ and the metal abundance $Z$.  The correlation is then extrapolated to zero metallicity to estimate the primordial mass fraction of helium $Y_{\rm P}$. Since the universe was born with no heavy elements, $Y_{\rm P}=Y(Z=0)$. This method has been modified to use the O abundance instead of $Z$, because measuring other chemical elements becomes impractical. Although some groups seek alternatives to O, like using N or S \citep[e.g.][]{Pagel1992, Vital2018}.

\section{Observations}

We analyzed the \ion{H}{II} region NGC~346 of the Small Magellanic Cloud.  We obtained the long-slit spectra with FORS1, at the VLT located at Melipal, Chile.  We used IRAF to reduce the spectra, following the standard procedure. We also corrected for underlying absorption and reddening correction, as well as a renormalization of H$\beta$. For further information please see \citet{Valerdi2019}.

\section{Physical Conditions}

We used optical collisional excitation lines (CELs) to calculated physical conditions, which were determined fitting the line intensities with {\texttt{PyNeb}} by \citet{Luridiana2003}. The temperature and density  were determined using the conventional sets of intensity ratios from the auroral to the nebular lines. Electron temperatures were determined using [\ion{O}{III}] $\lambda4363/(\lambda4959+\lambda5007)$, [\ion{O}{II}] ($\lambda7320+\lambda7330$)/($\lambda3726+\lambda3729$), and [\ion{N}{II}] $\lambda5755/\lambda6584$. Electronic densities were determined from the [\ion{O}{II}] $\lambda3726/\lambda3729$, and [\ion{S}{II}] $\lambda6731/\lambda6716$ ratios, which are strongly density dependent. In addition, we obtained the [\ion{Fe}{III}] density from the computations by \citet{Keenan2001}, using the $I(\lambda4986)/I(\lambda4658)$ ratio. The temperatures and densities are presented in Table \ref{tab:tem-cel}. 

\begin{table}[htb]
\centering
\caption{Temperature and density from CELs. \label{tab:tem-cel}}
\begin{tabular}{lcccc}
\hline
\hline
{Temperature (K)} & [OIII] & [OII] & [NII] & HeI \\
NGC~346 & 12871$\pm$98 & 12445$\pm$464 & 10882$\pm$767 & 11400$\pm$550\\ \hline
Density (cm$^{-3}$) & [OII]	& [SII]	& [FeIII] & HeI\\
NGC~346 & 23.7$\pm8.2$ & 32$\pm$21	& $101\pm17$ & $<$10	\\ 
\hline
\end{tabular}
\end{table}

We also obtain values for the temperature and density from \ion{He}{I} lines. We used the program {\tt Helio14}, which is an extension of the maximum likelihood method presented by \citet{Peimbert2000} to obtain these values. By using ten \ion{He}{I} lines, we obtained a temperature $T_{e}$(\ion{He}{I}) and a density $n_{e}$(\ion{He}{I}).

\section{Chemical Abundances}

We obtained ionic abundances using both: the direct method and the $t^2$ formalism. We first used {\texttt{PyNeb}} to perform the computations $t^2=0.00$. To obtain the ionic abundance using $t^2\neq0.00$, we used the equation 11 by \citet{Peimbert2004}. To obtain the $t^{2}$ value, we combining results from CELs and values from \ion{He}{I} and using the {\tt Helio14} code we found that the maximum likelihood values are: $t^{2}$(He$^{+})=0.033\pm0.017$ and $T_0=11900\pm450$ K. The ionic abundance of He$^{++}$ was obtained from the $\lambda4686$ line and using the recombination coefficients given by \cite{Storey1995}. The Table \ref{tab:ionic} shows the ionic abundances for $t^{2}=0.00$ as well as for $t^{2}=0.033\pm0.017$.

\begin{table}[htb]
\centering
\caption{Ionic abundances for the \ion{H}{II} region NGC~346. \label{tab:ionic}}
\begin{tabular}{lcc}
\hline
\hline
Ion  & $t^{2}=0.000$  & $t^{2}=0.033\pm0.017$ \\ 
\hline
He$^{+}$	& 10.917$\pm$0.004	& 10.915$\pm$0.004	\\
He$^{++}$	& 8.30$\pm$0.04		& 8.30$\pm$0.04		\\
\hline
N$^{+}$		& 5.87$\pm$0.02	& 5.97$\pm$0.06	\\
O$^{+}$		& 7.41$\pm$0.05	& 7.54$\pm$0.08	\\
O$^{++}$	& 7.93$\pm$0.01	& 8.04$\pm$0.06	\\
Ne$^{++}$	& 7.22$\pm$0.01	& 7.33$\pm$0.06	\\
S$^{+}$		& 5.55$\pm$0.03	& 5.65$\pm$0.06	\\
S$^{++}$	& 6.18$\pm$0.02	& 6.24$\pm$0.04	\\
Ar$^{++}$	& 5.54$\pm$0.01	& 5.60$\pm$0.03	\\
Ar$^{+3}$	& 5.11$\pm$0.02	& 5.22$\pm$0.06	\\
Cl$^{++}$	& 4.37$\pm$0.03	& 4.47$\pm$0.06	\\
\hline
\multicolumn{3}{c}{In units of $12+\log n(X^{+i})/n({\rm H})$.}
\end{tabular}
\end{table}

To establish the total abundance of an element, one must account for the unobserved ionization stages, by means of introducing an Ionization Corrector Factor (ICF). The functional forms of those ICFs frequently depend on the depth, quality, and coverage of the spectra. 

For \ion{H}{II} regions with a high degree of ionization, we can consider that the presence of neutral helium within is negligible, and ${\rm ICF}({\rm He}^0)\simeq1$. To obtain the abundance of helium we used: 
\begin{equation}
\frac{N({\rm He})}{N({\rm H})} = {\rm ICF}({\rm He}^{0})\left[
\frac{N({\rm He}^{+})+N({\rm He}^{++})}{N({\rm H}^{+})}\right],
\end{equation}
previous computations by \citet{Luridiana2003}, \cite{Peimbert2002}, \cite{Peimbert2007}, and \cite{Relano2002} showed that the ICF(He$^0$)$=1.00$ for NGC~346.

\begin{table}[htb]
\centering
\caption{Total abundances for the \ion{H}{II} region NGC~346. \label{tab:tot-cel}} 
\begin{tabular}{ccc}
\hline
\hline 
Element  & $t^{2}=0.000$  & $t^{2}=0.033\pm0.017$ \\ 
\hline
He	& 10.918$\pm$0.004  & 10.916$\pm$0.004 	\\
N	& 6.50$\pm$0.03	&	6.61$\pm$0.07 \\
O	& 8.05$\pm$0.02	&	8.19$\pm$0.08 \\
Ne	& 7.33$\pm$0.02	&	7.48$\pm$0.08 \\
S	& 6.40$\pm$0.03	&	6.44$\pm$0.08 \\
Ar	& 5.70$\pm$0.02	&	5.82$\pm$0.07 \\
Cl	& 5.37$\pm$0.04	&	5.47$\pm$0.07 \\
\hline
\multicolumn{3}{c}{In units of $12+\log n(X)/n({\rm H})$.}
\end{tabular}
\end{table}

For oxygen, to obtain the total abundance, we considered that $N({\rm O}^{3+})/N({\rm O})= N({\rm He}^{++})/N({\rm He})$ \citep{Peimbert1969}, therefore:
\begin{equation}
\frac{N({\rm O})}{N({\rm H})} = {\rm ICF}({\rm O}^{+}+{\rm O}^{++})\left[\frac{N({\rm O}^{+})+N({\rm O}^{++})}{N({\rm H}^{+})}\right],    
\end{equation}
where 
\begin{equation}
{\rm ICF}({\rm O}^{+}+{\rm O}^{++}) = \frac{N({\rm O}^{+})+N({\rm O}^{++})+N({\rm O}^{3+})}{N({\rm O}^{+})+N({\rm O}^{++})}.    
\end{equation}

For nitrogen, and neon we obtained the total abundances using the ICFs by \citet{Peimbert1969}. We observed the auroral lines of [\ion{S}{II}] and [\ion{S}{III}]. To obtain the total sulphur abundance we used the ICF proposed by \citet{Stasinska1978}. For the case of argon, we have measurements for lines of [\ion{Ar}{III}] and [\ion{Ar}{IV}]; we used the calculations by \citet{PerezM2007}. Finally for chlorine, we have measurements for lines of [\ion{Cl}{III}], and to obtain the total abundance we used the ICF proposed by \citet{Delgado2014}. In Table \ref{tab:tot-cel} we present the results for $t^{2}=0.00$ as well as for $t^{2}=0.033\pm0.017$.

\section{Primordial helium abundance}

Through the chemical abundances of NGC~346, we computed the primordial helium abundance. To obtain this value, we need to calculate the fraction of helium by mass. For low metallicity objects such as NGC~346, its oxygen is expected to represent $O/Z\approx0.55$ of the total mass of heavy elements \citep{Peimbert2007}. In this way, the chemical composition of NGC~346 is $X=0.7465$, $Y=0.2505$ and $Z=0.0030$.

To compute $Y_{\rm P}$, we assumed that:
\begin{equation}
Y_{\rm P}=Y-O\phantom{1}\frac{\Delta Y}{\Delta O},    
\end{equation}
and we adopted $\Delta Y/\Delta O=3.3\pm0.7$ \citep{Peimbert2016}, this value derived from metal-poor extragalactic \ion{H}{II} regions with different O and $Y$ values.

Through the low-metallicity \ion{H}{II} region  NGC~346 we have determined $Y_{\rm P}$. By measuring its \ion{He}{1} line intensities, we determined a primordial helium abundance $Y_{\rm P}=0.2451\pm0.0026$. Our determination is consistent with literature values, but our error bar is smaller. For further discussion on the determination of this value, and its error bars please see \citet{Valerdi2019}.

Table \ref{tab:helium}, shows $Y_{\rm P}$ measurements reported in the literature. We note that our result is consistent with the results by \citet{Aver2015}, \citet{Peimbert2016}, and \citet{Vital2018}. However we note that the \citet{Izotov2014} determination differs significantly from the other values. We consider that they have some systematic errors that they are ignoring; for example, part of this difference is due to the adopted temperature structure.

\begin{table}[htb]
\centering
\caption{$Y_{\rm P}$ values basen on \ion{H}{II} regions. \label{tab:helium}} 
\begin{tabular}{lc}
\hline
\hline
$Y_{\rm P}$ source & $Y_{\rm P}$  \\ 
\hline 
\citet{Izotov2014}    &  $0.2551\pm0.0022$ \\
\citet{Aver2015}      &  $0.2449\pm0.0040$ \\
\citet{Peimbert2016}  &  $0.2446\pm0.0029$  \\
\citet{Vital2018}     &  $0.245\pm0.007$    \\
This work &  $0.2451\pm0.0026$ \\
\hline
\end{tabular}
\end{table}

Through our determination of $Y_{\rm P}$, as a complement, we estimated the number of neutrino families and neutron half-life value. Using the $Y_{\rm P}$ value reported, and assuming a neutron half-life value $\tau_{n}= 880.2$ s \citep{Data2018}, we obtained that $N_{\nu}=2.92\pm0.20$. As expected our result is consistent with  \citet{Aver2015}, \citet{Peimbert2016}, and \citet{Vital2018}; these results agree with the presence of three families of neutrinos, which is consistent with laboratory determinations \citep[e.g.,][]{Mangano2005, Mangano2011}. While the $Y_{\rm P}$ result by \citet{Izotov2014} suggests the presence of a fourth neutrino family (see Table \ref{tab:parameters}).

\begin{table}[htb]
\centering
\caption{Predicted equivalent number of neutrino families, $N_{\nu}$ and the neutron half-life, $\tau_{\nu}$. \label{tab:parameters}} 
\begin{tabular}{lccc}
\hline
\hline
$Y_{\rm P}$ source & $Y_{\rm P}$ & $^{a}N_{\nu}$ & $^{b}\tau_{\nu}$ (s)   \\ 
\hline 
\citet{Izotov2014} & $0.2551\pm0.0022$ & $3.58\pm0.16$ & $921\pm11$ \\
\citet{Aver2015} & $0.2449\pm0.0040$ & $2.91\pm0.30$ & $872\pm19$   \\
\citet{Peimbert2016} & $0.2446\pm0.0029$ & $2.89\pm0.22$ & $870\pm14$ \\
\citet{Vital2018} & $0.245\pm0.007$   & $2.92\pm0.55$ & $872\pm33$   \\
This work & $0.2451\pm0.0026$ & $2.92\pm0.20$ & $873\pm13$           \\
\hline
\multicolumn{4}{l}{\tablenotemark{a} Assuming $\tau_{\nu}= 880.2$ s by \citet{Data2018}.} \\
\multicolumn{4}{l}{\tablenotemark{b} Assuming $N_{\nu}= 3.046$ by \citet{Mangano2005}.}
\end{tabular}
\end{table}

We also estimated the value of the neutron half-life using the value of $N_{\nu}= 3.046$ \citep{Mangano2005}, and we obtained $\tau_{n}=873\pm13$ s. Our result together with the results obtained from \citet{Aver2015}, \citet{Peimbert2016}, \citet{Vital2018}, are within 1$\sigma$ from the value presented by \citet{Data2018}, and they are in agreement with the $\tau_{n}$ laboratory determinations (see Table \ref{tab:parameters}). However, the results obtained by \citet{Izotov2014} differs by more than 3$\sigma$ from the laboratory determination.


\bibliographystyle{aaabib}
\bibliography{references}

\end{document}